\documentclass[preprint]{aastex}
 
\shorttitle{The composite SNR G326.3-1.8}
\shortauthors{Dickel, Milne, \& Strom}

\begin{document}
 
 
\title{Radio Emission from the Composite Supernova Remnant G326.3$-$1.8
  (MSH~15-5$\it 6$)}

 
\author{John R. Dickel\altaffilmark{1}}
\affil{Astronomy Department, University of Illinois at Urbana-Champaign,
    Urbana IL 61801}
\email{johnd@astro.uiuc.edu}
 
\author{D. K. Milne}
\affil{Australia Telescope National Facility, Epping NSW 1710, Australia}
\email{dmilne@atnf.csiro.au}
 
\and
 
\author{Richard G. Strom\altaffilmark{2}}
\affil{ASTRON, 7990AA Dwingeloo, The Netherlands}
\email{strom@nfra.nl}
 
 
\altaffiltext{1}{visiting astronomer at ASTRON, 7990AA Dwingeloo, The
  Netherlands} 
 
\altaffiltext{2}{ also at Astronomical Institute, University of
  Amsterdam, The Netherlands} 
 
\begin{abstract}
High resolution radio observations of the composite supernova remnant
(SNR) G326.3$-$1.8 or MSH 15-5${\it 6}$ with the Australia Telescope
Compact Array show details of both the shell and the bright plerion
which is offset about 1/3 of the distance from the center of the SNR to
the shell.  The shell appears to be composed of thin filaments, typical
of older shell SNRs.  The central part of the elongated plerion is
composed of a bundle of parallel ridges which bulge out at the ends and
form a distinct ring structure on the northwestern end.  The magnetic
field with a strength of order 45 ${\mu}$Gauss, is directed along the
axis of the ridges but circles around the northwestern ring.  This
plerion is large and bright in the radio but is not detected in
x-ray or optical wavelengths.  There is, however, a faint hard x-ray
feature closer to the shell outside the plerion.  Perhaps if the
supernova explosion left a rapidly moving magnetar with large energy
input but initially rapid decay of both relativistic particles and
magnetic field, the observed differences with wavelength could be explained.
\end{abstract}
 

 
\section{Introduction}

The supernova remnant (SNR) G326.3$-$1.8 or MSH15-5${\it 6}$
\citep{m61} is the prototype of the class of
composite SNRs containing a shell with a relatively steep radio
spectrum and an interior flat-spectrum plerion component \citep{w88}.  
In some composite SNRs a pulsar is observed in the
plerion, e.g. 0540$-$693 in the Large Magellanic Cloud \citep{s84,m93} 
but in a number of others, including
G326.3-1.8, no pulsar has been detected.  In all cases, however, a
pulsar is presumed to be present to inject high energy particles into the
plerion component.  The lack of detection is attributed to a
misalignment of the pulsar's axis so that its beam does not cross the
earth.  Alternatively, the pulsar may be very old and faint so that
more sensitive surveys are needed to detect it or the pulsed emission
may have ceased altogether.  

The shell, on the other hand, is produced by the interaction of
supersonic ejecta from the explosion and swept-up material from the
surroundings.  Strong shocks heat the gas to produce x-rays and
accelerate particles to relativistic speeds which, together with
magnetic fields amplified in the interaction zone, produce synchrotron
radiation.  Subsequent cooling and compression of the shocked gas can
locally increase the synchrotron emission and also produce optical
recombination-line emission.

There is a great range of relative sizes and brightnesses of the
plerion and shell components in composite remnants.  The large size
and high brightness of the plerion in G326.3$-$1.8 place it near the
extreme in each category.  It fills about 0.04 of the area and has
about 0.20 of the integrated flux density of the entire remnant.  By
contrast, the well known SNR W44 represents the other end of the range
for comparison: its plerion covers only about 0.0025 of the area and
has a flux density of 0.001 that of the entire SNR \citep{g97,f96}.  

To understand the relation between the different components and how
they might evolve, we need detailed studies of both the shell and the
plerion in several composite SNRs covering the full range of
parameters from G326.3$-$1.8 to W44.  G326.3$-$1.8 has been investigated at
a number of wavelengths.  The highest resolution radio images to date
are from the MOST at 0.843 GHz \citep{k87,m89,w96} and from the Fleurs
Synthesis Telescope at 1.414 GHz \citep{m89};  both had half-power
beamwidths of about 45 arcsec.  These show the faint, rather incomplete
shell but were unable to resolve fine scale features within the
plerion.  The shell is slightly non-circular with a diameter
between about 40 and 35 arcmin.  The plerion is displaced about 7
arcmin to the southeast from the center of the SNR;  its outline is
irregular but is contained within an ellipse having axes of 10 arcmin
$\times$ 5 arcmin.  The long axis is aligned approximately SE-NW.  The 
best polarimetry available, by \citet{m89}, has a resolution of 3
arcmin at 8.4 GHz.  The plerion is strongly polarized with the magnetic 
field aligned approximately along its major axis.  

The x-ray emission also
shows shell structure in some locations \citep{k93}.  Extended
features in both hard and soft x-rays appear within the interior
although they show no correspondence to the radio plerion nor to other
features seen in the radio or optical \citep{p98}.  Optical
filaments are present around parts of the shell and on the face of the
SNR \citep{vdb79,z79} but, again, show little
relation to any radio features.  The filaments show a high ratio of 
[S II]/H$\alpha$ which is typical of shocked material in radiative
equilibrium \citep{d80}.

Although the plerion is unseen in the other wavelength ranges, it
is clearly an important part of the SNR.  Its radio synchrotron
emission is strong, implying a large content of low-energy relativistic
electrons but apparently a rapid loss of high energy ones.
Determination of structure within the plerion may help establish the
relation to any pulsar.  Other characteristics can give indications of
its relation to other parts of the SNR and its evolution within it.  In 
addition, it is important to do a sensitive search for any radio
counterparts to the various x-ray features seen in the remnant.  

We adopt the kinematic distance determined from an H$\alpha$ radial
veocity measurement by \citet{r96} of 4.1 kpc.  This value is compatible
with a visual extinction measurement of $>$ 5.1 magnitudes by \citet{d80} and
a 21-cm absorption measurement of $>$ 1.5 kpc by \citet{cas75}.  At this
distance, the outer edge of the shell of G326.3$-$1.8 has dimensions of
48 $\times$ 42 pc and the extent of the irregular boundary of the
plerion is about 12 $\times$ 6 pc. 

To investigate the radio structure and polarization in
detail we have observed G326.3$-$1.8 with the Australia Telescope
Compact Array (ATCA).  Polarimetric images of the plerion and its
near environs have been obtained at 4.8 and 8.64 GHz while the entire
SNR has been imaged at 1.34 GHz.  These observations reveal the detailed
structure of the plerion and its polarization for the first time.  The
equipment and observations are described in section 2, the results are
presented in section 3, the interpretation is discussed in section 4,
and a summary is is section 5.

\section{Equipment and Observations}

The ATCA \citep{a92} can record both 1.34 and 2.4 GHz simultaneously or 
both 4.8
and 8.64 GHz simultaneously.  Both pairs of frequencies were observed
but the data at 2.4 GHz were seriously degraded by interference and not
used in the analysis.  Each set of observations was done with four separate
antenna configurations giving a total of 54 independent baselines
covering a range of spacings from 31 to 6000 meters.  The observing
parameters are given in Table 1.  The beams at the two higher
frequencies were nearly round and so were restored with circular
half-power beamwidths but the more elliptical beam at 1.34 GHz with a
HPBW of 8.5$''$ $\times$ 6.4$''$ at a position angle of 11$^{\circ}$, was left
with its original elliptical dimensions.

The large angular size and low surface brightness of G326.3$-$1.8
required long-integration, full-synthesis mosaic observations.  At 1.34
GHz, the entire SNR was covered by a 16-point mosaic pattern with
pointings spaced by 11 arcmin.  At 4.8 and 8.64 GHz it was decided to
concentrate only on the bright plerion component which was covered by a
25-point mosaic with spacings of 2.5 arcmin which was just under the
half width of the primary beam pattern at 8.64 GHz.  

The observing bandwidth was 128 MHz divided into 32 4-MHz channels.  At
4.8 and 8.64 GHz, edge channels were removed and the central 24
channels averaged together to give a single channel for the analysis.
At 1.34 GHz the central channels were maintained separately to
remove phase smearing near the edges of the fields and to prevent
possible Faraday depolarization across the band.  The data reduction
was done with the AIPS and MIRIAD packages.  After editing and
calibration were carried out, the mosaic images were formed using uniform
weighting and deconvolution was performed with CLEAN.  The mosaicing operation
automatically corrects for the primary beam response function.  The
total intensity image at 1.34 GHz was made with the multi-frequency
synthesis technique.  In all cases the missing short spacings and the
presence of bright nearby sources made confusion noise greater than the
thermal system noise.  This was particularly true for the total
intensities whereas the polarized intensities were weak enough that the
theoretical noise levels were almost reached.  The noise values listed
in Table 1 were the measured values on the images. 

For the polarization analysis, the images in the individual Stokes
parameters Q and U were deconvolved separately and then combined to
produce polarized intensity and position angle maps.  The intensities
were corrected for the Ricean bias which arises from the quadratic
combination of two intensities.  At 1.34 GHz, the initial data were
combined into five 20-MHz channels to produce five polarized intensity
and position angle maps.  This sufficiently reduced the Faraday
depolarization and also allowed good determination of the Faraday
rotation between these channels across the full band so that no
180$^{\circ}$ ambiguities in position angle were present.  The
polarized intensity maps for the individual channels were subsequently
averaged to produce a single map.  At the higher frequencies, the Faraday
depolarization effects are less significant and the entire bandwidth
was used for a single map at each frequency.

Flux density calibration was carried out using observations of PKS
B1934$-$638.  The antenna gains and polarization were calibrated by
regular observations of PKS B1520$-$58 throughout the observing runs.

\section{Results}

\subsection{Total Intensity}

\subsubsection{Morphology}

Figure 1 is a greyscale image of the whole SNR at a frequency of 1.34
GHz with a half-power beamwidth of 8.5$''$ $\times$ 6.4$''$.  The
entire periphery of the remnant can be seen as a faint complete shell.
This shell was not as obvious in previous lower-resolution images
\citep{m79,w96} which contained the full complement of short spacings
between elements and showed the emission across the full face of the
SNR.  The overall emission appears to be more filled in toward the 
southwest of the plerion than it is around the rest of the remnant.
The overall east-west alignment of filaments recognized in the previous
images is also present in this ATCA image, primarily on the eastern side.
Some of the individual filaments, particularly in the northwestern
corner, appear unresolved but often sit on a smoother background which
may contain several overlapping features. There are a number of faint
optical filaments in G326.3$-$1.8 but they do not form a complete shell
and several sit on the face of the remnant \citep{vdb79,z79,d80,r96}
where no particular radio enhancements are visible.      

The new image also reveals structure within the plerion for the first
time.  This component of the SNR appears virtually identical at all
three observing frequencies (see Figure 2 for the 4.8 and 8.64 GHz
images).  It consists of several parallel ridges running along the
major axis with some irregular patches at the ends.  In particular, a
circular ring is seen on the northwestern end.  There is smooth
emission between the ridges and over the interior of the ring.  All of
the individual features in the plerion appear to be resolved with half-power
thicknesses of about 15 - 20 arcsec.

\subsubsection{Spectrum}

As indicated above, it is impossible to measure the total flux density
of the whole SNR even at 1.34 GHz because of the missing short spacings
of the telescope array.  The plerion, however, is smaller and we
have tried to estimate its integrated flux density at all three
frequencies.  The major difficulty remains the determination of the
background level including emission from the shell component.  By
examination of various slices and images, we estimate the uncertainty
to be about 33$\%$ of the values determined. These three values,
together with all other published flux densities for both the plerion
and for the entire SNR are listed in Table 2 and shown in Figure 3.
Except for the few uncertainties listed, no other authors have been
brave enough to give values for their uncertainties. We have assumed
that they are all comparable and so have fit the spectra with
unweighted least-square lines. 

The resultant spectra with power-law slopes of $-$0.29 for the whole SNR
and $-$0.18 for the plerion are shown by the solid lines.  The dashed
line with a slope of $-$0.34 represents the spectrum of the whole SNR
minus the plerion or that of the shell component alone.  Both
components have spectral indices consistent with their classifications
although that of the shell is on the flat side of their distribution,
similar to the well-known shell remnant IC443 \citep{g98}.  The
integrated flux density of the shell at 1 GHZ is 114 Jy which
corresponds to a surface brightness of 1.23 $\times$ 10$^{-20}$ W m$^{-2}$
Hz$^{-1}$ sr$^{-1}$.  The plerion with a flux density of 26 Jy at 1 GHz
has a surface brightness of 7.8 $\times$ 10$^{-20}$ W m$^{-2}$ Hz$^{-1}$
sr$^{-1}$.  Both values are about average for objects of their
classification and size. 

\subsection{Polarization}

The intensity of the shell at 1.34 GHz is too low to evaluate the
polarization in detail but in Figure 4, we show a greyscale image of
the polarized intensity with superimposed total-intensity contours.
The plerion is bright and has a mean fractional polarization of 12\%.
Much of the apparent ring around the plerion in both total and
polarized intensities is a grating response but the polarized emission
is stronger on the shell and does fill in somewhat southwest of
the plerion toward the shell, just as the total intensity does.  The
subsequent discussion will concentrate on the plerion which is
significantly polarized at all three wavelengths.  Because the
resolution was about a factor of 2 better at the two higher
frequencies, we will concentrate on those results.  

Figure 5 shows the electric vectors of the polarized intensity at 4.8 GHz with
superimposed total-intensity contours.  For visibility on this and all
other polarimetric maps, the vectors are spaced by 12 arcsec or over
3 half-power beamwidths.  This spacing also means that each vector is
entirely independent.  It can be seen that the polarized intensity
follows the total intensity quite closely and Figure 6 is a greyscale
image of the fractional polarization at 4.8 GHz with the same
superimposed total-intensity contours.  The mean fractional
polarization is 35\% with no significant variation across the plerion.
The results at 8.4 GHz are very similar with a mean fractional
polarization of 40\%.  These fractional values are very high for
SNRs;  they are likely artificially increased in the aperture synthesis
observations because the polarized intensity probably has more fine
structure than does the total intensity and the lack of short spacings
between the antennas will misrepresent the background level under the
plerion.  Observations with the Parkes 63-m telescope find 20\% at 8.4
GHz \citep{m89}, 12\% at 5 GHz \citep{m75,w71}, and 2\% at 2.7 GHz
\citep{m72} toward the peak of the plerion.  Because the low resolution
of the single dish observations would smear the structure within the
plerion, the true values probably lie between these extremes.  

The position angles at 4.8 and 8.34 MHz can be used to evaluate the
Faraday rotation.  The results, shown in Figure 7, with filled boxes
indicating positive values of rotation measure and open boxes
negative values.  The rotation measure is quite uniform at $-$300 rad
m$^{-2}$ with reliable values falling between $-$200 and $-$400 rad
m$^{-2}$.  The few positive values seen are probably not real as they
all lie near the edges of the image where the polarized intensity is
too weak to determine the rotation measure accurately.  

Because the measured position angles can always have 180$^\circ$
ambiguities, there can be n $\times$ 180$^\circ$ additional rotation -
either plus or minus - between the two frequencies.  The 1.34 GHz
results for the 5 adjacent 20-MHz bands, however, also give $-$300 rad
m$^{-2}$ for the rotation measure and thus confirm the adopted result.

The rotation measure can be used, in turn, to derotate the measured
position angles to their zero-wavelength or intrinsic values.  Figure 8
shows the actual direction of the ${\it {magnetic}}$ field vectors of
the plerion in G326.3$-$1.8.  The lengths are proportional to the
polarized intensity at 4.8 GHz and the contours are, again, the total
intensity at 4.8 GHz.  We note that the field aligns beautifully with
the ridges in the total intensity and on the northwestern end of the
plerion the field direction actually curves around to follow the ring
struture.

\section{Discussion}

G326.3$-$1.8 is a composite SNR with an evolutionarily old faint and thin
shell but a bright plerion which may still be receiving input from an
unseen pulsar.  The shell looks like those which are well into
the point-blast stage of development in which the radius, R,
expands as R $\propto$ t$^{0.4}$ where t is time.  The outline may be
slightly flattened to the northwest but must be expanding fairly
uniformly.  The overall x-ray emission is fairly uniform with only
marginal indications of a shell structure but it does appear brighter
toward the center and southeastern parts of the remnant \citep{k93}.
Infrared data \citep{a89}, on the other hand, show the apparent
Galactic ridge to the northwest so the reduction of apparent x-ray
emission there could be absorption by intervening material which could
also be beginning to interact with the expanding shell. 

The lack of correlation of the radio and x-ray emission in the interior
of this SNR is very unusual. The missing x-ray emission from the radio
plerion is particularly hard to understand.  It appears to be the only known
plerion with strong radio emission but no x-rays.  We note that, if
anything, there is an indication of a decrease in the infrared
structure at the position of the plerion \citep{a89}.  If radio
emission trails behind a pulsar the brightest part of the radio plerion
could be behind the hardest x-rays.  Such a situation appears to be
true for the plerion N157B in the LMC \citep{l00} but there is still
significant radio emission from the suspected position of the x-ray pulsar
\citep{m98,w98} and significant x-ray emission from the position of the
brightest radio emission in that object.  A pulsar moving at
perhaps 500 km sec$^{-1}$ could have covered the approximate 8 pc
distance from the center of the SNR to the radio plerion in about
15,000 years and then have moved outward to the position of the current
hard x-ray feature in about another 15,000 years, leaving the plerion
behind. 

Estimation of the expected x-ray emission from the radio plerion
requires knowledge of the spectrum of the emission between the x-ray
and radio wavelengths. Theoretical predictions \citep{p73,r84}, with
some observational support \citep{s89,g92}, indicate that the
synchrotron emission generated by a pulsar should have a break in its
spectrum from synchrotron radiation losses.  This will move downward in
frequency with time as the highest energy particles lose their energy
the fastest.  To determine the break frequency and thus the overall
spectrum, we need to know the magnetic field strength.  Using equations
5.7 to 5.12 of \citet{g65} for synchrotron
emission we find that the emission is proportional to the product of
the relativistic electron energy and the magnetic field strength to the
1 - $\alpha$ power, where $\alpha$ is the spectral index.  Adopting the
observed spectral index of -0.18 and the 1 GHz flux density of 26 Jy
found above and then integrating the radio emission from 10 MHz to 100
GHz for the plerion at a distance of 4.1 kpc, we find an integrated
luminosity of 3 $\times$ 10$^{34}$ erg sec$^{-1}$.  The total energy 
density in relativistic electrons radiating between 10 MHz and 100 GHz
is then 9 $\times$ 10$^{42}$/H$^{1.18}$ ergs.  The three dimensional
volume of the plerion is, of course, unknown but we shall approximate
it as a cylinder with a length of 12 pc and a radius of 3 pc.  The
energy density is then 1.8 $\times$ 10$^{-15}$/H$^{1.18}$ erg
cm$^{-3}$.  In Kepler's SNR \citep{m84} and N23 in the LMC \citep{d98},
the approximate ratio of relativistic electron energy to magnetic
energy is about 3, so we will adopt this value to arrive at a mean
magnetic field strength of 45 $\mu$Gauss.  This value is high for an
old shell SNR but low for a young plerion where particle injection and
magnetic field amplification are still occurring.  It may indeed be
appropriate for an older plerion.  Such a field would give a break
frequency near 1.6 $\times$ 10$^{13}$ Hz for a lifetime of 3 $\times$
10$^{4}$ years. For this spectral break, the integrated 0.5-10 keV
x-ray luminosity of the radio plerion should be about 2 $\times$
10$^{38}$ erg sec$^{-1}$, orders of magnitude above that detected
toward the hard x-ray feature to the west.  Even with large
uncertainties in the parameter estimates, this must indeed be a very 
unusual source not to be detected in x-rays.  
 
The extended feature in the ASCA hard x-ray band (2.5 - 9 keV) between
the plerion and the shell to the southwest \citep{p98} is 
intriguing as it might represent the current location of the pulsar
presumed to be responsible for the various emission components.
Although there is a slight increase in the overall radio emission in
that general area, it appears as if it is a brightness enhancement on
the face of the shell.  There are also several optical filaments in
that area \citep{vdb79,d80}, but there is no morphological association 
of any feature with the x-rays.  The integrated x-ray emission is so 
faint, about 3 $\times$ 10$^{33}$ erg sec$^{-1}$ between 0.5 and 10 keV
\citep{p98}, however, that it would be missed by the radio surveys.
Although we cannot measure the field in that direction, if the break
frequency is the
same as toward the radio plerion, the expected radio flux density 
at 1.34 GHz would be of order 1/2 mJy from the whole 6 $\times$ 3 arcmin
area of the hard x-ray source.  Even with the 45-arcsec beam of the
MOST, such emission would not be detectable.  We note that a stronger
magnetic field as might be expected near the pulsar, would move the
break frequency lower and increase the expected radio emission but the
time scale near the pulsar is also shorter which will work in the
opposite direction to reduce the ratio of x-ray and radio emission. 

One possible way to make G326.3$-$1.8 the necessary  extreme case of
particle injection and decay might be for it to have been a magnetar
with an extremely high initial magnetic field so that it very quickly
gave up most of its energy and stopped producing the hardest x-rays.
The current production of x-ray and radio emitting particles is very
low but the lower energy electrons from the initial much more energetic
pulsar would have persisted longer and, as indicated above, they can
still be radiating significantly after some thousands of years provided
that the average magnetic field strength does not much exceed the
current value of 45 $\mu$Gauss estimated above.    

The structure of the plerion is unusual wherever the pulsar may be
located.  The morphology shows a bundle of parallel ridges running from
southeast to northwest.  Presumably injection from a pulsar could
create one tube or perhaps a spiral pattern, e.g. SS433 \citep{h81}, but
it is difficult to see how to get the aligned structures, much less the
ring with a diameter of about 3.6 pc at one end.  The motion of any pulsar
relative to the observed structures also remains unclear.  If it
started in the center of the remnant and moved outward to the present
position of the hard x-ray component, the nearly perpendicular plerion
coming out from one side near the middle of the path is very peculiar.
If the ring structure represents a bubble around the
original position of the pulsar and the pulsar then moved along the
axis of the plerion, it had to make an abrupt turn to get to the
present location of the hard x-rays.  Detection of a pulsar and
measurements of its proper motion would greatly aid in understanding
this peculiar object.     

The magnetic field is also aligned along the ridges but wraps
around with the ring structure on the northwestern end (Figure 8).
The relatively high fractional polarization and very uniform Faraday
rotation suggest that there is little internal Faraday effect and that
the magnetic field is quite uniform.  The field is probably frozen into
the fluid structure.  Unfortunately, it is not possible to determine
the thermal energy and pressure to see if that dominates the magnetic
field in the plerion since it is seen only in radio synchrotron emission.

\section{Summary}

As often happens in astronomy, the first identified object
in the class of composite SNRs is one of the most unusual.  While the
shell component appears quite normal, the plerion has several unique
characteristics.  It is certainly one of the largest and brightest
plerions relative to its shell although there are several apparently
naked plerions which are larger including G328.4+0.2 \citep{g00}, N157B
\citep{c92,l00} with a detected 16 msec x-ray pulsar \citep{m98},
and G74.9+1.2 \citep{w78} all of which reach a diameter of about 25
pc.  

While most plerions are elongated like that in G326.3$-$1.8, they
generally show a rather amorphous or irregular structure rather
than the apparent bundle of parallel filaments seen in this object.
Further, the ring on the northwestern end might be a preexisting
bubble or somehow related to the pulsar and plerion.  The magnetic
field strength is moderate and the field follows the direction of the
ridges but also circles around the ring.

There is no x-ray emission specifically associated with the very bright
radio plerion.  Perhaps if the stellar remains of the supernova
explosion were a magnetar with a magnetic field strength of
$\sim$ 10$^{15}$ Gauss and rapid energy loss, the initial emission would have
been very bright but have dropped quickly.  The break frequency in the 
synchrotron spectrum could also have dropped quickly but then would
have slowed as the plerion expanded and the magnetic field strength
dropped.  This process could reduce the x-ray emission quickly but
leave significant radio emission.  The very faint hard x-ray emission
to the southwest of the radio plerion could represent what is left of
the stimulation by the magnetar.

\acknowledgements
We thank Paul Plucinsky for information on the x-ray emission.  Brian
O'Shea helped with some of the data analysis.  JRD acknowledges a
Visitor's Fellowship from the Netherlands Organization for Scientific
Research (NWO) during his very enjoyable stay at ASTRON.

\clearpage

\begin{deluxetable}{lcccc}
\tablecolumns{5}
\tablecaption{Observing parameters}
\tablehead{
\colhead{Frequency (GHz)} & \colhead{HPBW} & \multicolumn{2}{c}{rms
  Noise (mJy beam$^{-1}$)} & \colhead{Flux Density of}  \\
\cline{3-4}
\colhead{} & {} & \colhead{Total Intensity} & \colhead{Polarized
  Intensity} & \colhead{PKS B1934$-$638 (Jy)}}
\startdata
~~1.34 & 8.5$''$ $\times$ 6.4$''$\tablenotemark{a} & 1.70 & 0.13 & 15.01 \\
~~4.80 & 3.8$''$ $\times$ 3.8$''$ & 1.30 & 0.05 & ~5.83 \\
~~8.64 & 3.2$''$ $\times$ 3.2$''$ & 0.80 & 0.10 & ~2.84 \\
\enddata
\tablenotetext{a}{long axis at position angle 11$^{\circ}$}
\end{deluxetable}

\clearpage

\begin{deluxetable}{lrrl}
\tablecolumns{4}
\tablecaption{Flux Densities of G326.3-1.8}
\tablehead{
\colhead{Frequency (GHz)} & \multicolumn{2}{c}{Flux Density (mJy)} &
\colhead{Reference} \\
\cline{2-3} \\
\colhead{} & \colhead{Entire SNR} & \colhead{Plerion} & \colhead{}}
\startdata
~~~0.408 & 180~~ & ~ & ~~~Clark et al. 1975 \\
~~~0.843 & 153 $\pm$ 40~~ & ~ & ~~~Milne et al. 1989 \\
~~~0.843 & $>$ 130~~ & 22~~~~~~ & ~~~Whiteoak \& Green 1996 \\
~~~1.34 & $\gg$ 60~~ & 30~~~~~~ & ~~~this paper \\
~~~1.4 & $>$ 95~~ & ~ & ~~~Milne et al. 1979 \\
~~~2.65 & 115~~ & ~ & ~~~Milne 1972 \\
~~~4.8 & ~ & 25~~~~~~ & ~~~this paper \\
~~~5.0 & 98~~ & ~ & ~~~Milne 1969 \\
~~~5.0 & 75~~ & ~ & ~~~Milne \& Dickel 1975 \\
~~~8.4 & 68~~ & ~ & ~~~Milne et al. 1989 \\
~~~8.64 & ~ & 15~~~~~~ & ~~~this paper \\
~~~8.8 & ~ & $<$ 40~~~~~~ & ~~~Dickel et al. 1973 \\
~~14.7 & 69 $\pm$ 8~~ & 16 $\pm$ 7~~~~~~ & ~~~Milne et al. 1979 \\
\enddata
\end{deluxetable}

\clearpage

\figcaption[f1.eps,f1b.eps]{a)  Image of the composite supernova remnant
  G326.3$-$1.8 at a frequency of 1.34 GHz with a half-power beamwidth
  of 8.5$''$ $\times$ 6.4$''$. b)  A copy of the MOST image at 0.843
  GHz with a resolution of 45$''$ for comparison. \label{Fig. 1 a,b}}

\figcaption[f2a.eps,f2b.eps]{Images of the plerion component of
  the supernova remnant G326.3$-$1.8 at a) 4.8 GHz with a circular
  half-power beamwidth of 3.8$''$ and b) 8.64 GHz with a circular
  half-power beamwidth of 3.4$''$. \label{Fig. 2 a,b}} 

\figcaption[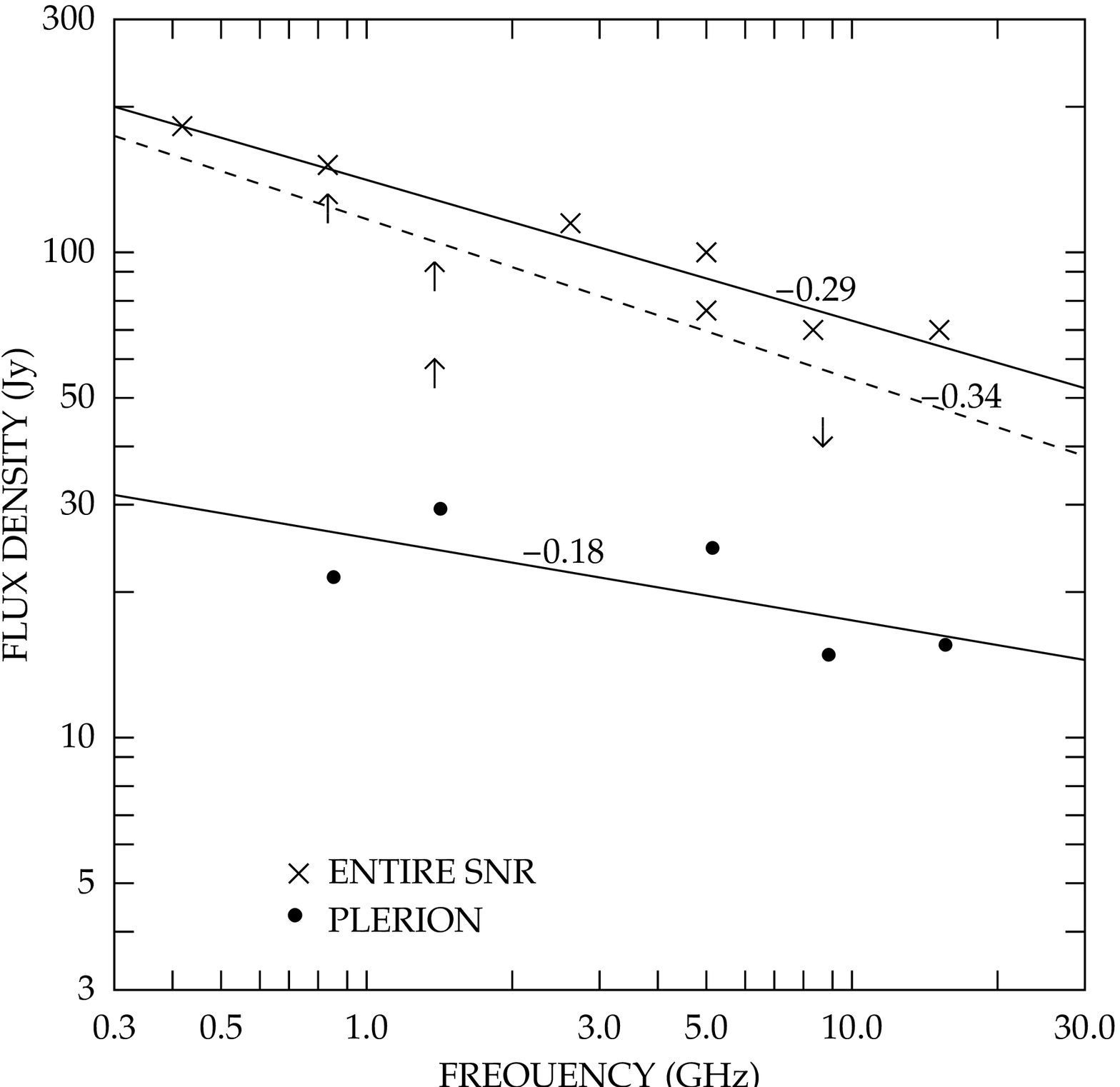]{The radio spectrum of G326.3$-$1.8 from the
  data in Table 2.  The $\times$s represent the entire SNR and the dots
  represent only the plerion component.  The up and down arrows
  indicate upper and lower limits, respectively.  The upper line is the
  fit for the whole SNR;  the lower one is for the plerion;  and the middle
  dashed line is the difference spectrum representing the shell part of
  this composite SNR. \label{Fig. 3}}  

\figcaption[f4.eps]{Greyscale of the polarized intensity of the
  supernova remnant G326.3$-$1.8 at a frequency of 1.34 GHz with selected
  superimposed contours of total intensity. \label{Fig. 4}} 

\figcaption[f5.eps]{Electric vectors of the polarized emission
  from the plerion in the supernova remnant G326.3$-$1.8 at a frequency
  of 4.8 GHz with total intensity contours.  A vector length of 4
  arcsec represents a brightness of 0.525 mJy beam$^{-1}$.  The
  half-power beamwidth is shown by the dot within the box in the lower
  right corner of the map. \label{Fig. 5}}

\figcaption[f6.eps]{Greyscale of the fractional polarization of
  the plerion in the supernova remnant G326.3$-$1.8 at a frequency of
  4.8 GHz with total intensity contours. \label{Fig. 6}}

\figcaption[f7.eps]{Faraday rotation measure toward the region of
  the plerion in the supernova remnant G326.3$-$1.8 with total
  intensity contours at a frequency of 4.8 GHz.  The filled boxes
  represent positive values of rotation measure and the open boxes
  negative values.  A box dimension of 4 arcsec represents a rotation
  measure of 200 rad m$^{-2}$. \label{Fig. 7}} 

\clearpage

\figcaption[f8.eps]{Vectors representing the direction of the
  magnetic field in the region of the plerion in the supernova remnant
  G326.3$-$1.8.  The vector length represents the polarized intensity
  at 4.8 GHz and the greyscale is the total intensity at 4.8
  GHz. \label{Fig. 8}}

\end{document}